\documentclass[a4paper]{jpconf}
\usepackage{graphicx}
\usepackage{orcidlink}
\usepackage{wrapfig}

\usepackage{lineno}
\usepackage[square,sort&compress,numbers]{natbib}

\begin{document}
\title{Comparing and improving hybrid deep learning algorithms for identifying and locating primary vertices}

\author{S Akar$^1$\,\orcidlink{0000-0003-0288-9694}, M Peters$^1$, H Schreiner$^2$, M D Sokoloff$^1$\,\orcidlink{0000-0001-6181-4583}, W Tepe$^1$}

\address{$^1$ University of Cincinnati, Cincinnati, OH 45221, USA}
\address{$^2$ Princeton University, Princeton, NJ 08544, USA}

\ead{simon.akar@cern.ch}

\begin{abstract}
 Using deep neural networks to identify and locate proton-proton collision points,
 or primary vertices, in LHCb has been studied for several years. 
 Preliminary results demonstrated the ability for a hybrid deep learning algorithm 
 to achieve similar or better physics performances compared to standard heuristic 
 approaches. 
 The previously studied architectures relied directly on 
 hand-calculated  Kernel Density Estimators (KDEs) as input features. 
 Calculating these KDEs was slow, making use of the DNN inference 
 engines in the experiment's real-time analysis
 (trigger) system problematic.
Here we present recent results from a high-performance hybrid deep learning algorithm that uses track parameters as input features rather than KDEs, opening the path to deployment in the real-time trigger system. 
\end{abstract}
\vskip -0.2in
\vspace{-0.2in}
\section{Introduction}
\label{intro}
The LHCb experiment was recently upgraded for Run 3 of the LHC. 
It will record proton-proton collision data at five times the instantaneous luminosity of Run 2. 
The average number of visible primary vertices (PVs), proton-proton collisions in the detector closest to the beam-crossing region, will increase from 1.1 to 5.6. 
The experiment will move to a pure software data ingestion and trigger system, eliminating the Level 0 hardware trigger altogether~\cite{LHCbCollaboration:2014vzo}. 
A conventional PV finding algorithm~\cite{Reiss:2749592, Reiss:CtD2020} that satisfies all requirements defined in the Trigger Technical Design Report~\cite{Aaij:2018jht} serves as the baseline. 
In parallel, we have been developing a hybrid machine learning algorithm, designed to run in the initial stage of the LHCb upgrade trigger.

The initial algorithm defined a one-dimensional Kernel Density Estimator (KDE) histogram, plus two more one-dimensional histograms, to describe the probabilities of tracks traversing small voxels in space.
These feature sets were modified to use two KDEs rather than one.~\cite{Fang:2019wsd,akar2020updated,Akar:2021gns}.
``Classic" convolutional Neural Networks (CNNs), referred to as {\tt KDE-to-hists} models, produce one-dimensional histograms that nominally predicts Gaussian peaks at the locations of true PVs using three (or four) input histograms as their feature sets. 
A hand-written clustering algorithm identifies the candidate PVs and their positions. 
A first set of results used a ``toy Monte Carlo'' with proto-tracking~\cite{Fang:2019wsd}. 
Using track parameters produced by the LHCb Run~3 Vertex Locator (VELO) tracking algorithm~\cite{Hennequin:2019itm} leads to significantly better performance~\cite{akar2020updated}.

The original KDE~\cite{akar2020updated} is a projection of a three-dimensional probability distribution in voxels that has contributions {\em only} when two tracks pass close to each other. 
Calculating this KDE exactly is very time-consuming.  One of the goals
of our project is to predict PV positions directly from tracks' parameters.
A proof-of-concept was presented last year~\cite{Akar:2021gns}. 
We trained a model, {\tt tracks-to-KDE}, predicting the KDE using the tracks parameters as input and  merged it with a trained version of a {\tt KDE-to-hists} model. 
Mediocre performance, compared to the original {\tt KDE-to-hists} model, was 
achieved after a final training allowing all weights of the merged 
{\tt tracks-to-hists} model to float.

In the last PV-finder update~\cite{akar2020updated}, we also presented studies of alternative {\tt KDE-to-hists} model architectures. In particular, we showed that the capacity of our original CNN architecture~\cite{Fang:2019wsd} to learn increased with the number of parameters, as expected from first principles.
The study of a modified version of the popular U-Net architecture~\cite{ronneberger2015unet}, showed similar performances compared to the best classic CNN architectures, and it trained more quickly.
Here we present improved results from an updated architecture for {\tt tracks-to-hists}. 
A brief description of the model in given in Sec.~\ref{sec:model}.
Performance is discussed in Sec.~\ref{sec:perf}. A summary is presented in Sec.~\ref{sec:summary}

%%%%%%%%%%%%%%%%%%%%%%%%%%%%%%%%%%%%%%%%%%%%%%%%%%%
%%%%%%%%%%%%%%%%%%%%%%%%%%%%%%%%%%%%%%%%%%%%%%%%%%%
%%
%%           PERFORMANCE EVOLUTION 
%% 
%%%%%%%%%%%%%%%%%%%%%%%%%%%%%%%%%%%%%%%%%%%%%%%%%%%
%%%%%%%%%%%%%%%%%%%%%%%%%%%%%%%%%%%%%%%%%%%%%%%%%%%
\section{Model architecture and training strategy}
\label{sec:model}

%%%%%%%%%%%%%%
\begin{figure}[t]
\centering
\includegraphics[width=0.95\textwidth,clip]{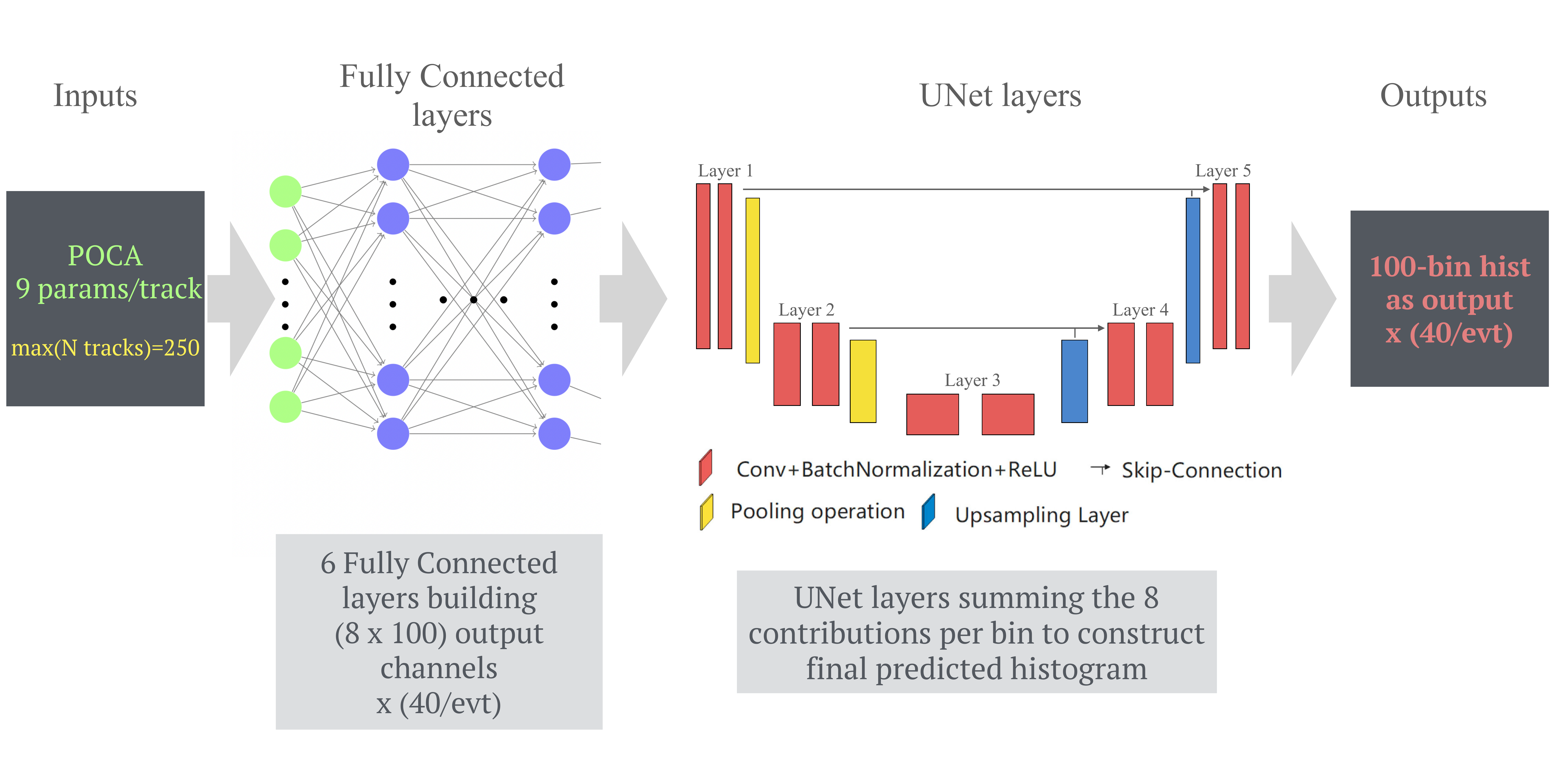}
\vspace{-5pt}
\caption{
This diagram illustrates the end-to-end, {\tt tracks-to-hist}, deep neural network used to predict an event's target histogram from its tracks' {\tt poca-ellipsoid}s.
Each event is now sliced in 40 independent 100-bin histograms.
Six fully connected layers populate 8 100-bin channels in the last of these layers, for each track. 
These contributions are summed and processed by a U-Net model with 5 convolutional layers to construct the final 100-bin histogram.
}
\label{fig:model}
\vskip -0.1in
\end{figure}
%%%%%%%%%%%%%%

Building on our previous experience, we define a merged {\tt tracks-to-hists} model; this architecture is based on the one used in the proof-of-concept presented in the last update~\cite{Akar:2021gns}.
The latest {\tt tracks-to-hists} model, whose architecture is shown in Fig.~\ref{fig:model}, includes a few updates: the {\tt tarcks-to-KDE} part of the model consists of 6 fully connected layers that are initially trained to
produce a KDE and the weights of the first 5 layers are temporarily
frozen; a variation with 8 latent feature sets is merged to the
trained {\tt KDE-to-hists} DNN where the classical CNN layers are 
replaced by a U-Net model.
Critically, we also updated the structure of the input data for training
and inference.   
 In the original approach, the one-dimensional feature sets consisted of 4000 bins along the z-direction (beamline), each $100\,{\rm \mu m}$ wide, spanning the active area of the VELO around the interaction point, such that $z \in [-100, 300]\,{\rm mm}$. 
In place of representing each event by a single 4000-bin histogram (feature set), 
we now 
slice each event into 40 slices of 100 bins each.
This approach is motivated by the fact that the shapes of the target histogram 
are expected to be invariant as a function of the true PV position
and it is easier for a DNN to learn to predict target histograms 
over a smaller range of bins.
Also, with an average of $\sim 5$ PVs per event, most of the bins in both the KDE and target histograms have no significant activity.
The 40 slices of 100 bins are independent and homogeneous between events. 
Each slice is treated independently, after which the predicted 4000-bin
histogram is stitched back together.

Training on nVidia RTX 2080Ti GPUs is performed in several steps. 
First, weights for the {\tt tracks-to-KDE} model that uses individual tracks parameters evaluated at their points of closest approach 
(poca-ellipsoid) to the beamline~\cite{Akar:2021gns} as input features to predict the KDE are determined.
Then,
the U-Net {\tt KDE-to-hists} model is trained using the 
hand-calculated KDEs 
as the input feature set.
Finally all weights and biases in the combined {\tt 
tracks-to-hists} model are allowed to float. 
An asymmetry parameter between the cost of overestimating contributions to the 
target histograms and underestimating them~\cite{Fang:2019wsd} is 
used as a hyperparameter to allow higher efficiency by incurring higher false
positive rates.

%%%%%%%%%%%%%%
\begin{figure}[t!]
\includegraphics[width=0.95\textwidth,clip]{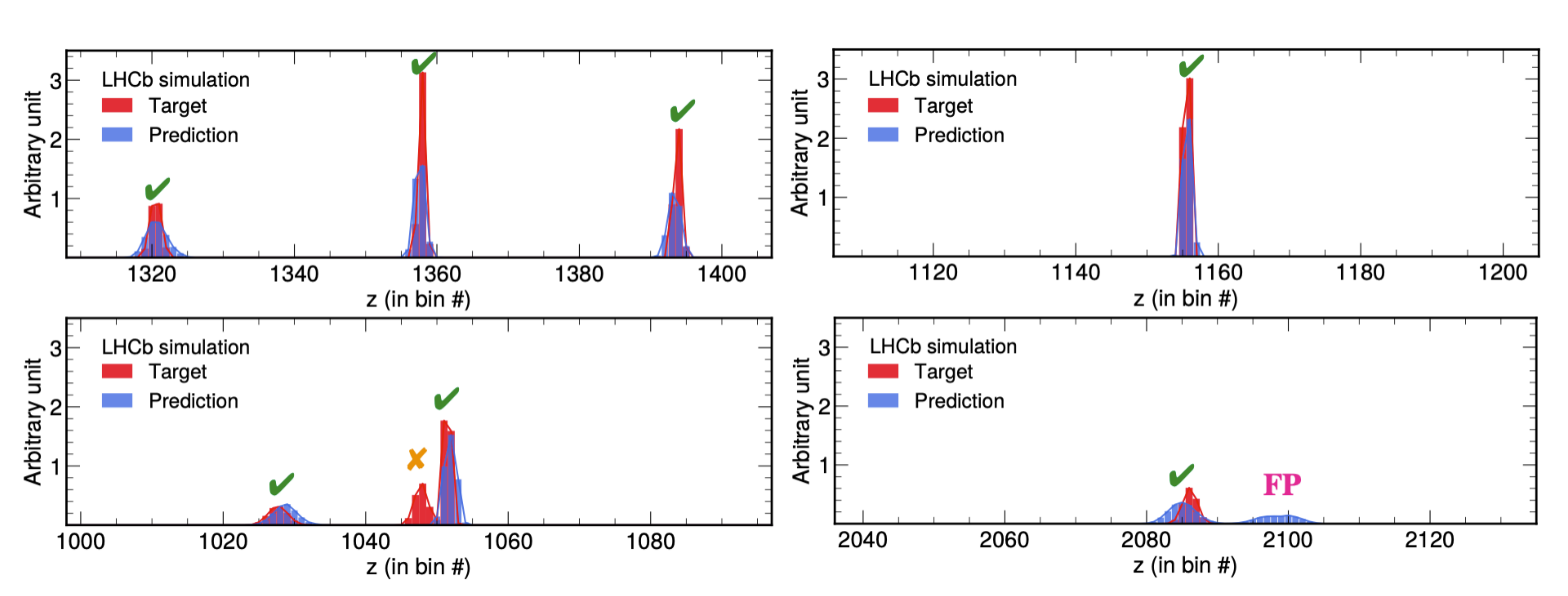}
\vspace{-10pt}
\caption{
Plots illustrating typical examples of target (shown in red) versus predicted (shown in blue) histogram.
Each of the four plots is drawn from different slices of different individual events.   
Each of the 100 bins in the histogram corresponds to a range in the $ z $-direction of $ 100 \,  \mu$m.
Predicted PVs that are matched to the correct true PV (majority of cases) are identified by a green tick in the plots. 
Example of missed true PV (orange cross label) and false positive prediction (FP label) are shown in the bottom-left and bottom-right plots, respectively. 
} %% end of \caption
\label{fig:targetHist}       % Give a unique label
\vskip -0.1in
\end{figure}
%%%%%%%%%%%%%%

%%%%%%%%%%%%%%%%%%%%%%%%%%%%%%%%%%%%%%%%%%%%%%%%%%%
%%%%%%%%%%%%%%%%%%%%%%%%%%%%%%%%%%%%%%%%%%%%%%%%%%%
%%
%%           PERFORMANCE EVOLUTION 
%% 
%%%%%%%%%%%%%%%%%%%%%%%%%%%%%%%%%%%%%%%%%%%%%%%%%%%
%%%%%%%%%%%%%%%%%%%%%%%%%%%%%%%%%%%%%%%%%%%%%%%%%%%
\section{Performances}
\label{sec:perf}

\paragraph{Evaluation}
Performance evaluation is obtained by a matching procedure done by a heuristic algorithm, based on the PV positions along the beam axis, $z$.
Figure~\ref{fig:targetHist} shows typical examples of predicted and target histograms using the final {\tt tracks-to-hists} model.
In the previous report of PV-finder performances~\cite{Fang:2019wsd,akar2020updated,Akar:2021gns} a predicted PV was matched if the distance, $\Delta z$, between it's position, $z_{\rm pred}$, and the true PV position, $z_{\rm true}$, satisfied $\Delta z = |z_{\rm pred}-z_{\rm true}| \leq 0.5 \,{\rm mm}$.
The false positive rate is obtained from the ratio of remaining predicted PVs after 
the matching procedure over the total number of true PVs.
This approach suffered a few limitations, independent of the intrinsic DNN 
algorithm performances.
For instance, it is known that a PVs resolution depends on the number of tracks 
originating from it. 
In case of low-multiplicity PVs, the predicted histogram from the DNN algorithm, 
similar in shape to the target histogram, can be displaced more than $0.5 \, {\rm 
mm}$ from the true PV position. 
This scenario can simultaneously ``produce" a missed PV (reduced 
efficiency) and a false positive signal (increased false positive rate).
To address this issue, a new matching procedure is used. 
Instead of a fixed window around the predicted PV position, a resolution function is introduced in the matching procedure. 
The resolution for a given predicted PV, $\sigma(z)$, is a function of the standard deviation of the predicted histogram multiplied by a constant parameter that can be tuned.
Using this definition, a predicted PV is now matched if $\Delta z  = |z_{\rm pred}-z_{\rm true}| \leq 5 \sigma(z)$. 
In Fig.~\ref{fig:targetHist} we observe that for matched PVs, the standard 
deviation of the predicted peaks (blue entries) is slightly larger than that
of the corresponding true PVs (red entries), estimated from the number of 
tracks.
Peaks with larger standard deviations are typically associated with PVs with fewer tracks.

\begin{figure}
\centering
\includegraphics[width=0.49\textwidth,clip]{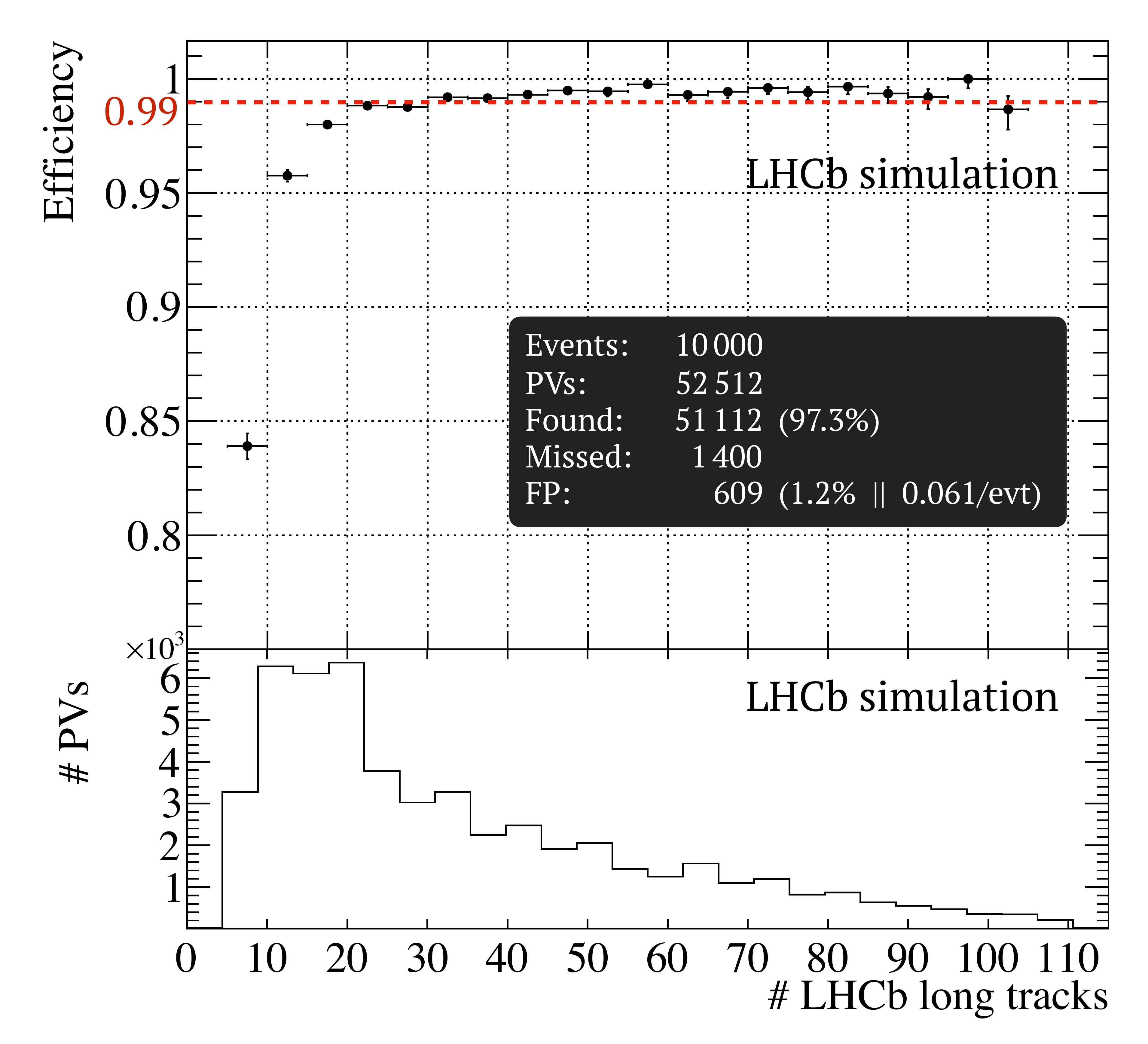}
\includegraphics[width=0.49\textwidth,clip]{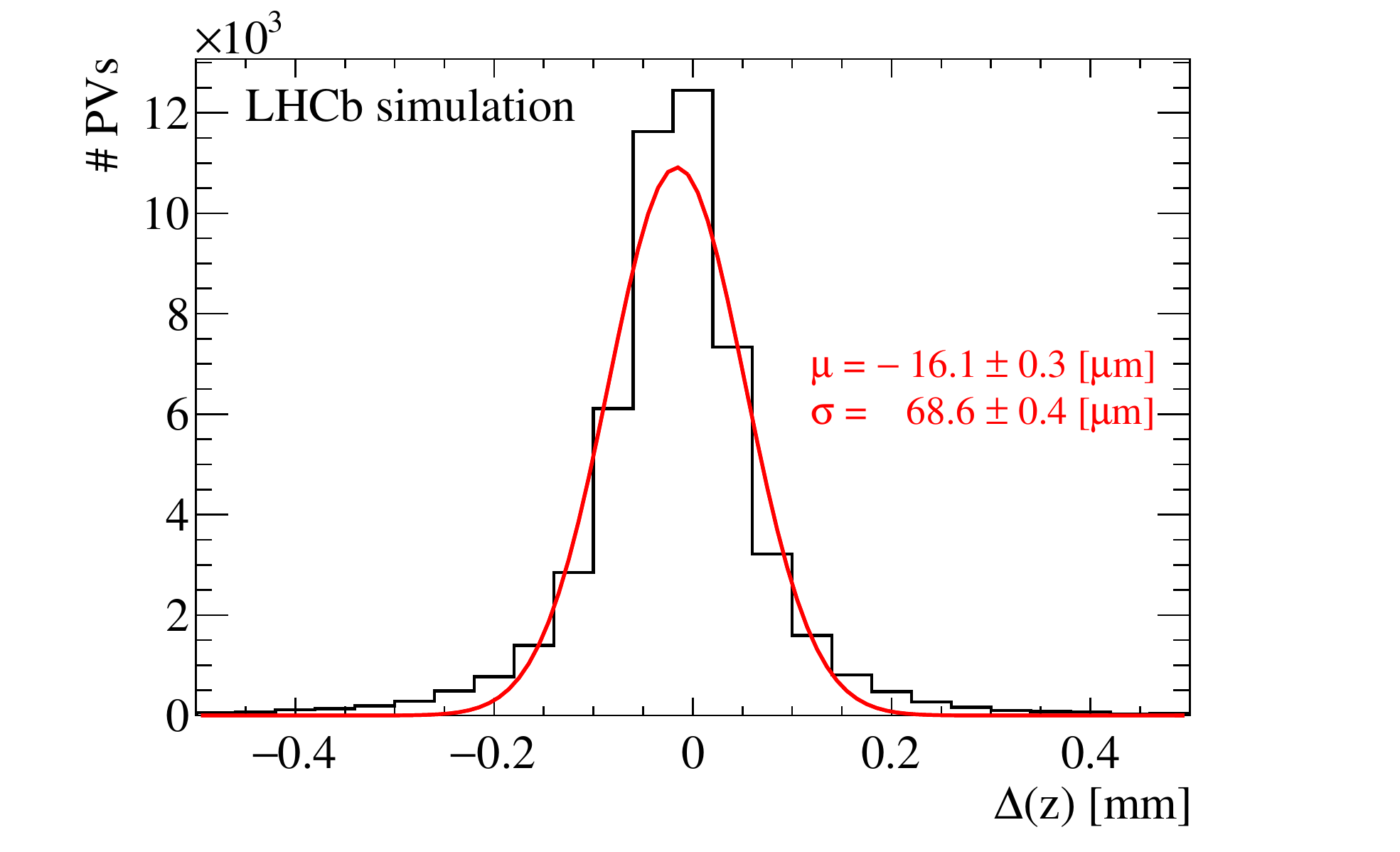}
\vspace{-5pt}
\caption{
(left) Efficiency of finding true PVs as a function of the number of fully reconstructed originating tracks. 
The integrated efficiency and false positive rate are also reported.
(right) Distribution of the distance, $\Delta z$, between the predicted and true PV position for matched PVs as defined in the text, overlaid by a fitted Gaussian.
}
\label{fig:eff_vs_nTracks}
\vskip -0.1in
\end{figure}

\paragraph{Predicted position and efficiency} 
The reported results come from statistically independent validation samples.
Using the matching procedure described above, we quantified the performances of the trained {\tt tracks-to-hists} DNN on a sample of 10K events corresponding to slightly more than 50K true PVs.
Figure~\ref{fig:eff_vs_nTracks} shows the efficiency of finding true PVs as a function of the number of fully reconstructed originating tracks, labelled LHCb long tracks, defined as tracks with hits in all the elements of the LHCb tracking system.
We observe that the efficiency is very high ($>99\%$) and stable for PVs with more than 20 tracks, and rapidly decreasing for PVs with fewer tracks. 
Also shown in Fig.~\ref{fig:eff_vs_nTracks} is the distribution of the distance, $\Delta z$, between the predicted and true PV position for matched PVs. 
From a fit using a Gaussian distribution, we observe a small bias of $-16.1\, \mathrm{\mu m}$ on the predicted PV position.
Most notably, the majority of matched PVs have a $\Delta z$ value below the bin width of the target histogram of $0.1 \, \mathrm{mm}$.  

%%%%%%%%%%%%%%
%%%%%%%%%%%%%%

%%%%%%%%%%%%%%
\begin{figure}
\centering
\includegraphics[width=0.65\textwidth,clip]{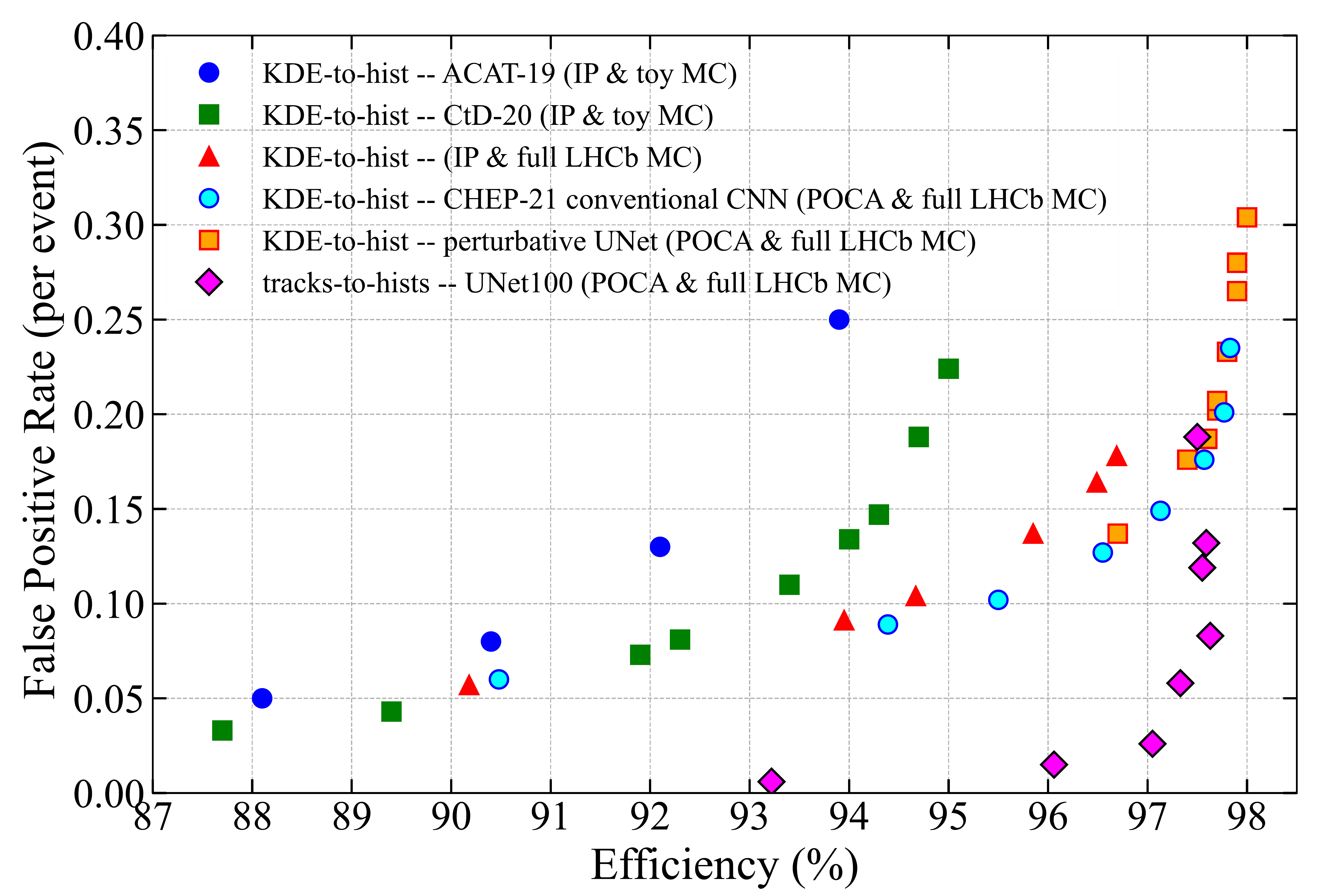}
\vspace{-5pt}
\caption{
Comparison between the performances of models reported in previous years and the newest model (magenta diamonds). An asymmetry parameter described in the text  is varied to produce the families of points observed.
}
\label{fig:evolution}
\vskip -0.1in
\end{figure}
%%%%%%%%%%%%%%

\paragraph{Performances evolution}
Figure~\ref{fig:evolution} shows how the performance of the DNN algorithms have evolved over time.
The efficiency is shown on the horizontal axis and the false positive rate per event is shown on the vertical axis.
The solid blue circles show the performance of any early {\tt KDE-to-hists} model described at ACAT-2019~\cite{Fang:2019wsd}.
The green squares show the performances of a {\tt KDE-to-hists} described at Connecting-the-Dots in 2020~\cite{akar2020updated}.
Both of the above models were trained using “toy Monte Carlo” with proto-tracking.
All subsequent DNN models were trained using full VELO tracking algorithm~\cite{Hennequin:2019itm}, leading to significantly better performances (red triangles to be compared to green squares). 
The cyan circles and the yellow squares correspond to the best achieved performances for {\tt KDE-to-hists} models using either a classical CNN architecture or a U-Net model described at CHEP-2021~\cite{Akar:2021gns}.
The performances of all above models were obtained using the ``old'' matching procedure with a fixed searching window of $0.5\,{\rm mm}$.
The magenta diamonds show the performance of the {\tt tracks-to-hist} model described in Sec.~\ref{sec:model}. 
These performances are obtained using the improved matching procedure described above.
The performance of the new {\tt tracks-to-hist} model enables the DNN
to simultaneously reach high 
efficiencies ($>97\%$) and low false positive rates ($0.03$ per event or $0.6\%$
per reconstructed PV).

%%%%%%%%%%%%%%%%%%%%%%%%%%%%%%%%%%%%%%%%%%%%%%%%%%%
%%%%%%%%%%%%%%%%%%%%%%%%%%%%%%%%%%%%%%%%%%%%%%%%%%%
%%
%%       Summary and Conclusions 
%% 
%%%%%%%%%%%%%%%%%%%%%%%%%%%%%%%%%%%%%%%%%%%%%%%%%%%
%%%%%%%%%%%%%%%%%%%%%%%%%%%%%%%%%%%%%%%%%%%%%%%%%%%
\section{Summary and Conclusions}
\label{sec:summary}

Since we presented results at CHEP 2021, we have updated the {\tt tracks-to-hists} 
model to produce the first competitive end-to-end hybrid deep learning algorithms 
for identifying and locating primary vertices based solely on tracks' parameters
as input features.
The performance of this model reaches very high efficiencies with low false positive rates.
By removing the (very slow) KDE calculation used as part of the hybrid {\tt KDE-to-hists} models, we have opened the path to include a PV finding DNN in the real-time trigger system of the experiment.
Studies are currently ongoing to validate the feasibility of including a version of the {\tt tracks-to-hists} model to execute in the CUDA cores (or in the tensor cores) of the GPUs constituting the LHCb trigger system.
We also plan on studying the effect of quantization on the model performances, as well as the size of the model itself (number of nodes).
Our specific algorithms are being designed for use in LHCb with its Run 3 detector.
Developing and deploying machine learning inference engines that are highly performant and satisfy computing system constraints requires sustained effort.
The results reported here should encourage work focused on using DNNs for identifying vertices in other high energy physics experiments as well.

%%%%%%%%%%%%%%%%%%%%%%%%%%%%%%%%%%%%%%%%%%%%%%%%%%%
%%%%%%%%%%%%%%%%%%%%%%%%%%%%%%%%%%%%%%%%%%%%%%%%%%%
%%
%%              Acknowledgments 
%% 
%%%%%%%%%%%%%%%%%%%%%%%%%%%%%%%%%%%%%%%%%%%%%%%%%%%
%%%%%%%%%%%%%%%%%%%%%%%%%%%%%%%%%%%%%%%%%%%%%%%%%%%
\section{Acknowledgments}
\noindent

The authors  thank the LHCb computing and simulation teams for their support and for producing the simulated LHCb samples used in this paper.
The authors  also thank the full LHCb Real Time Analysis team, especially the developers of the VELO tracking algorithm~\cite{Hennequin:2019itm} used to generate the ``full LHCb MC'' results presented in Fig~\ref{fig:evolution}.

This work was supported, by the U.S. National Science Foundation under Cooperative Agreement 
OAC-1836650 and awards 
PHY-1806260, 
and 
OAC-1450319.

%%%%%%%%%%%%%%%%%%%%%%%%%%%%%%%%%%%%%%%%%%%%%%%%%%%
%%%%%%%%%%%%%%%%%%%%%%%%%%%%%%%%%%%%%%%%%%%%%%%%%%%
%%
%%                BIBLIOGRAPHY 
%% 
%%%%%%%%%%%%%%%%%%%%%%%%%%%%%%%%%%%%%%%%%%%%%%%%%%%
%%%%%%%%%%%%%%%%%%%%%%%%%%%%%%%%%%%%%%%%%%%%%%%%%%%
%% mds 
%% mds For tables use syntax in table~\ref{tab-1}.
%% mds \begin{table}
%% mds \centering
%% mds \caption{Please write your table caption here}
%% mds \label{tab-1}       % Give a unique label
%% mds % For LaTeX tables you can use
%% mds \begin{tabular}{lll}
%% mds \hline
%% mds first & second & third  \\\hline
%% mds number & number & number \\
%% mds number & number & number \\\hline
%% mds \end{tabular}
%% mds % Or use
%% mds \vspace*{5cm}  % with the correct table height
%% mds \end{table}
%
% BibTeX or Biber users please use (the style is already called in the class, ensure that the "woc.bst" style is in your local directory)

\bibliography{ACAT22_PVFinder.bib}

\end{document}